\documentstyle[rotate,dina4,12pt,psfig]{article}
\newcommand{\be}{\begin{equation}}
\newcommand{\ee}{\end{equation}}
\newcommand{\bea}{\begin{eqnarray}}
\newcommand{\eea}{\end{eqnarray}}

\begin{document}
\title{%
A Dynamical Model of Color Confinement%
\thanks{Work supported by BMBF and GSI Darmstadt.}}
\author{%
S. Loh,
T.S. Bir\'o, U. Mosel and M.H. Thoma\\
Institut f\"ur Theoretische Physik, Universit\"at Giessen\\
D-35392 Giessen, Germany\\%
}
\maketitle
\begin{abstract}
A dynamical model of confinement based on a transport theoretical
description of the Friedberg-Lee model
is extended to explicit color degrees of freedom. The string tension is
reproduced by an adiabatic string formation from the nucleon ground state.
Color isovector oscillation modes
of a $q\bar{q}$-system are investigated for a wide range of relative
$q\bar{q}$-momenta and the dynamical impact of color confinement on the 
quark motion is shown.
\end{abstract}




\newpage

\begin{figure}
\caption{Expectation value of the $z$-coordinate $<z> = \int\int
z f^0(x,p) \; d^3x \; d^3p$ as a function of the time $t$  resulting
from an exitation energy of $3 GeV$ (dotted line).
For the solid line the oscillations have been averaged out.}
\label{rms}
\end{figure}
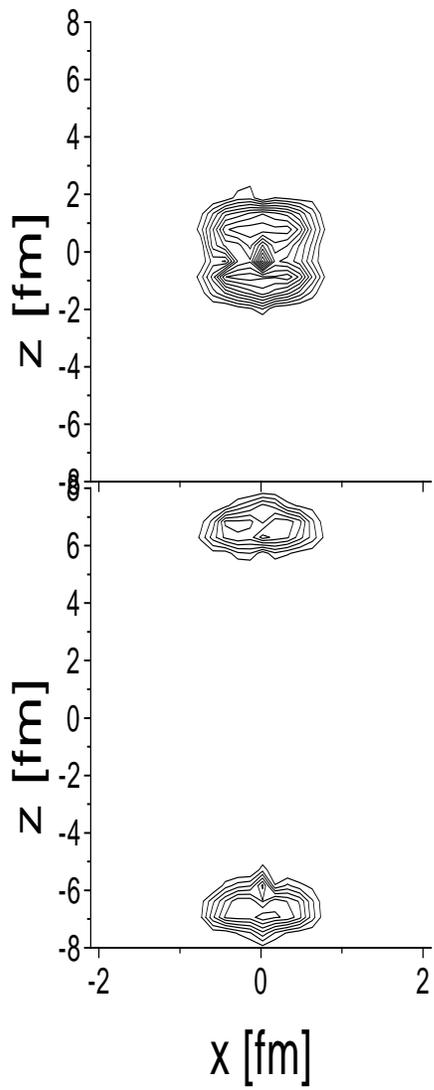
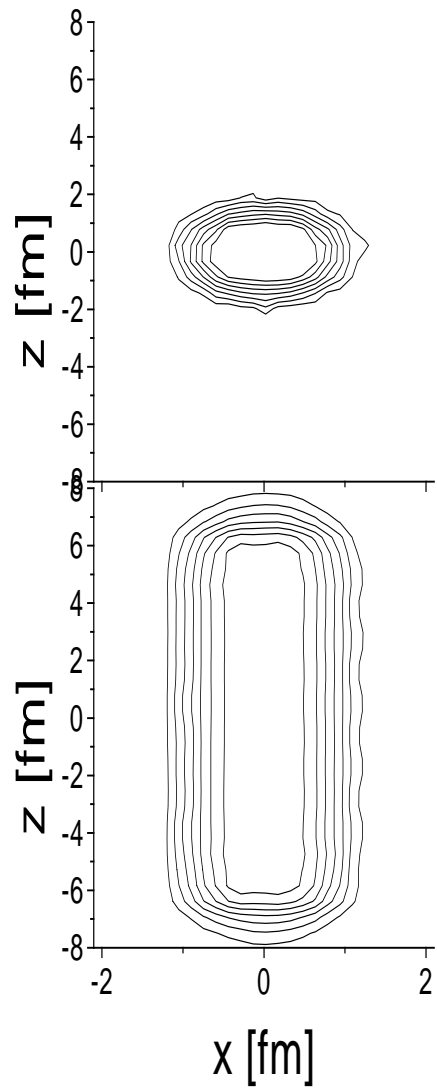
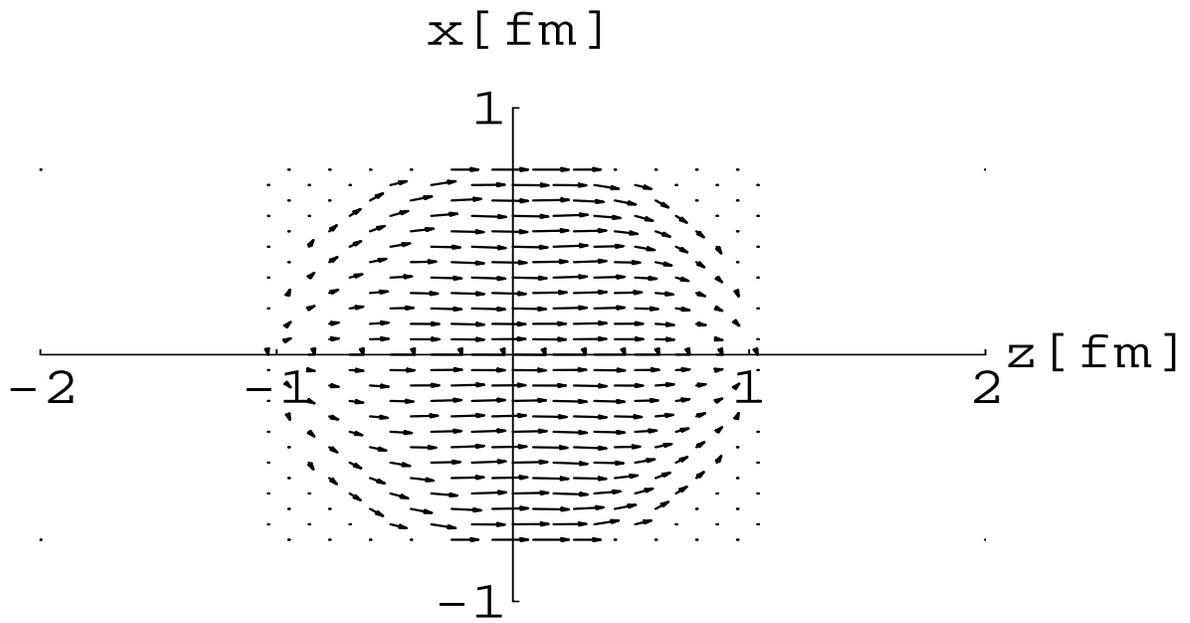
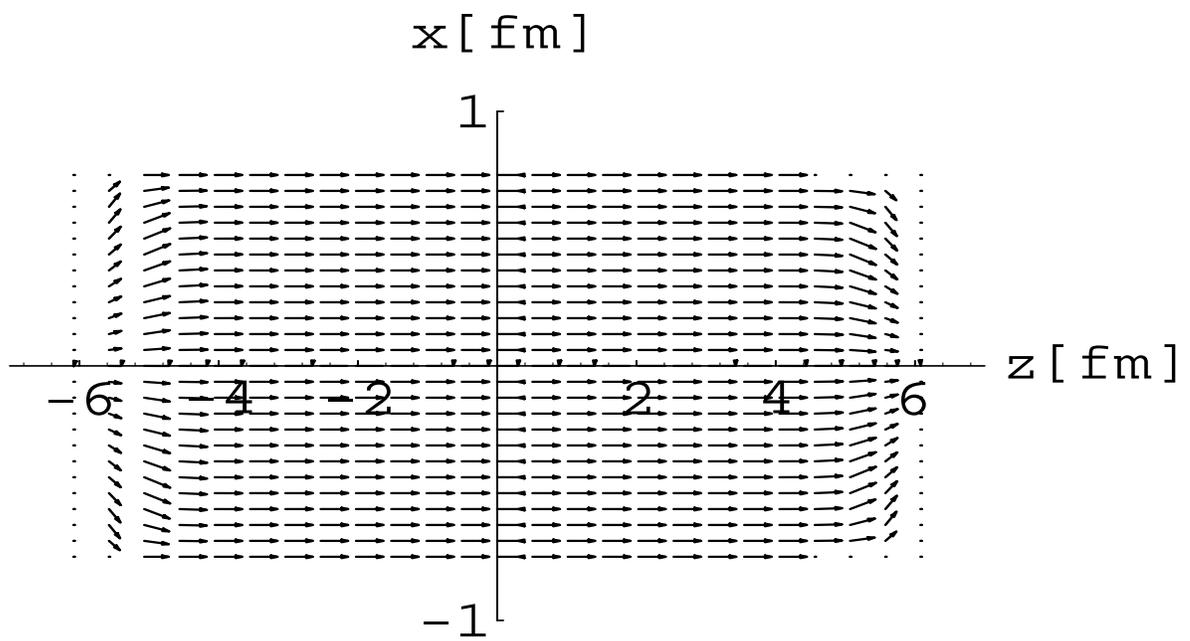
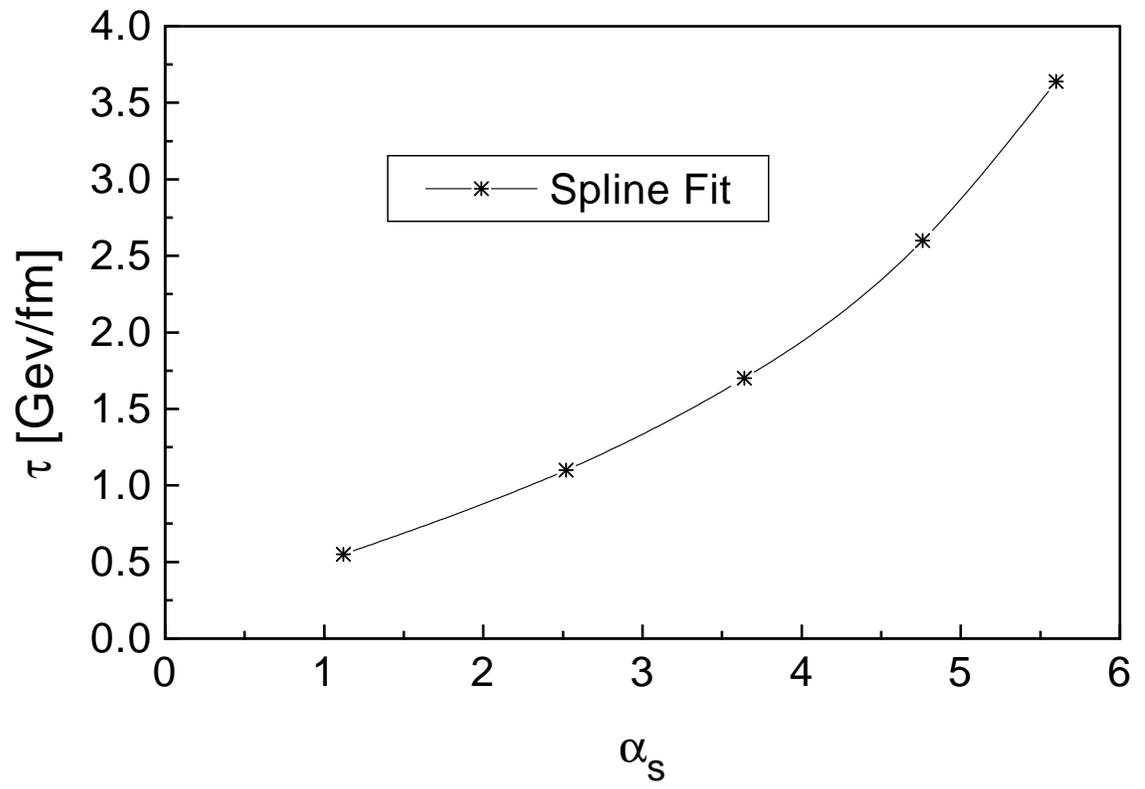
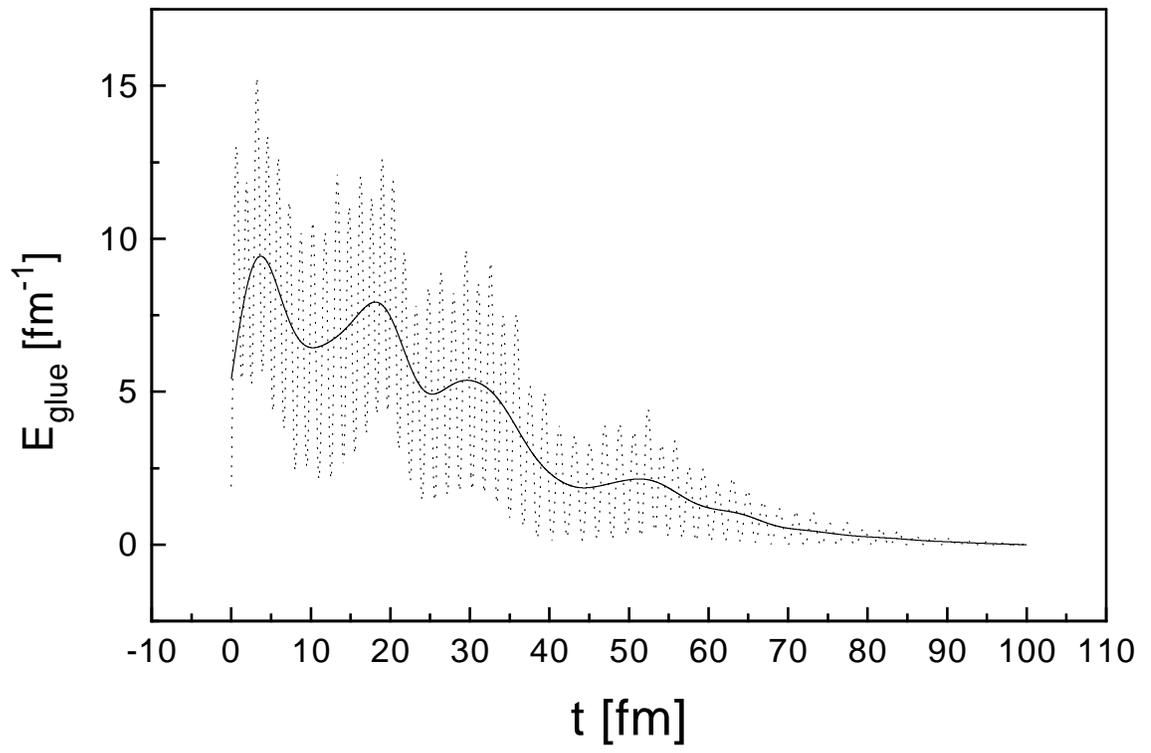
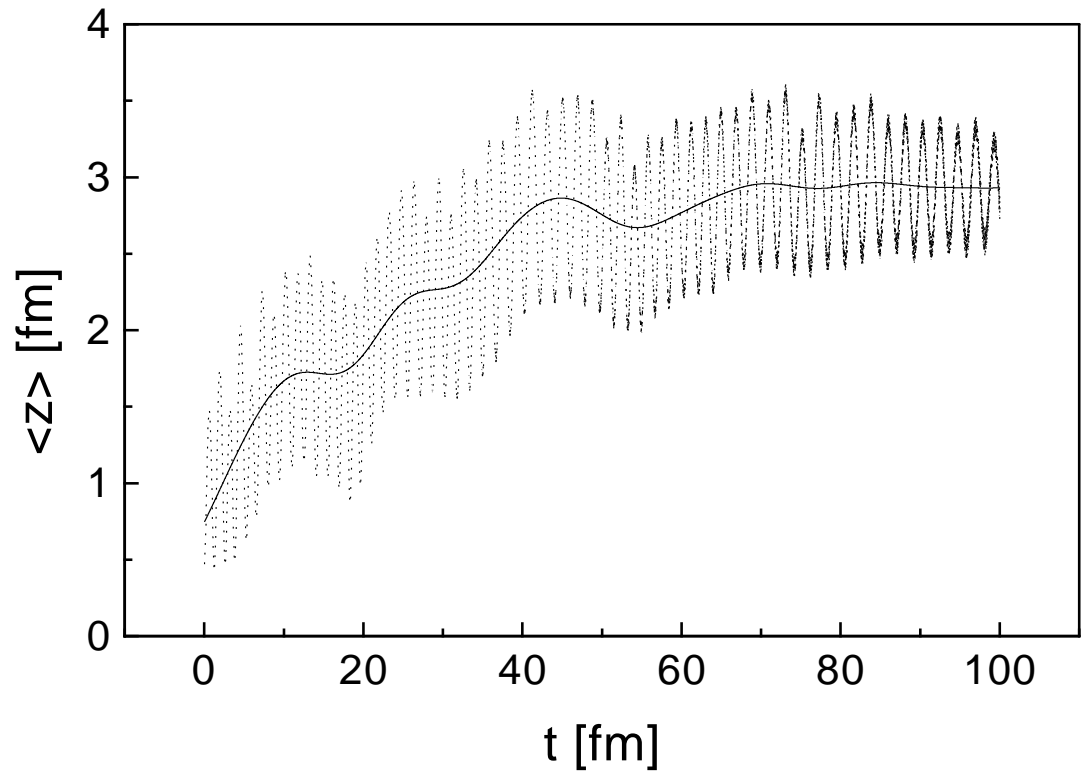

Kinetic theories are a well established tool in describing intermediate
energy heavy-ion collisions. Their success has stimulated the
development of transport theories based on QCD to describe collisions
for even higher energies.
For relativistic energies, string and
parton cascade models (Fritjof \cite{AnGuNi87}, Venus \cite{We89}, RQMD
\cite{SoStGr89/2}, HIJING \cite{WaGy91}, parton cascade \cite{GeMu92}) have
been developed. A serious problem of these models is the description of the
final state of the reaction, the hadronization, which can no longer be
treated perturbatively, since the hadrons are formed by the purely
nonperturbative effects of color confinement. Therefore the dynamics of
confinement should be considered in the transport descriptions.\\
There exist different approaches to deriving a transport theory that is
based on an effective quark model incorporating the dynamics of the
nonperturbative QCD vacuum.
Zhang and Wilets \cite{wil} have derived a transport theory based on the
Nambu Jona-Lasinio model in order to estimate chiral symmetry effects in
heavy-ion collisions; this is conceptually close to our work. However, besides
the absence of confinement, these authors do not actually perform dynamical 
simulations. Also soliton-soliton collisions in the framework of the 
Skyrme-model \cite{VeWaWaWy87},\cite{AlKoSeSo87} and of a
Skyrme-like $\sigma$-model \cite{KuPiZa93} have been
performed. However, these calculations do not include any explicit quark
degrees of freedom.\\
We assume that the nonperturbative effects can be modeled by the exchange 
of a scalar field, as it is successfully done in the Friedberg-Lee model
(for a review see \cite{wilbuch}). Since in this model the hadron surface
is generated dynamically, it is well suited for dynamical simulations.\\
In recent articles \cite{KaVeBiMo93,VeBiMo95} we presented the derivation
of the transport theory based on the Friedberg-Lee model
and applied it to simulations of nucleon-nucleon collisions for
bombarding energies of the order of a few GeV. In this work color
degrees of freedom were neglected.
Here we present the derivation and first dynamical applications
of a transport model {\em with color confinement}. After having fixed the strong
coupling constant with the help of the string constant, we will investigate
an excited state of the nucleon, the color isovector oscillation mode,
showing the dynamical impact of color confinement on the quark motion.\\
The phenomenological Lagrangian of the Friedberg-Lee model reads
\cite{FbLe77,FbLe78}
\be
  \label{lagrange}
  {\cal{L}} = \bar{\Psi}(i\gamma_{\mu}\partial^{\mu}-m_0-g_0\sigma)\Psi
              + {1 \over 2}(\partial_{\mu}\sigma)^2 - U(\sigma) 
              - {1 \over 4} \kappa(\sigma)F_{\mu\nu}^aF^{\mu\nu}_a 
         - ig_v\bar{\Psi}\gamma_{\mu}{\lambda^a \over 2}\Psi A^{\mu}_a \; ,
\ee
where $\Psi$ denotes the quark fields, $\sigma$ is the color singlet scalar
field representing the long range and nonabelian effects (of multi gluon
exchange), and the last term contains the interaction of the residual abelian
(one gluon exchange) color fields $A_{\mu}^a$
\be
  \label{ftensor}
  F_{\mu\nu}^a = \partial_{\mu}A_{\nu}^a - \partial_{\nu}A_{\mu}^a 
\ee
All the nonabelian effects are assumed to be absorbed in the color
dielectric function $\kappa(\sigma)$ which is chosen such that
$\kappa$ vanishes as $\sigma$ approaches its vacuum value $\sigma = \sigma_v$
outside the bag and $\kappa = 1$ inside.
$U(\sigma)$ is the self-interaction potential for the scalar
$\sigma$-field containing cubic and quartic terms.
The Euler-Lagrange equations derived from (\ref{lagrange}) are
\bea
  \label{dirac}
  (\gamma^{\mu}(i\partial_{\mu}-ig_v{\lambda_a \over 2}A_{\mu}^a)
  -m_0-g_0\sigma)\Psi = 0 \; , \\
  \label{klegor}
  \partial_{\mu}\partial^{\mu}\sigma + U'(\sigma) + {1 \over 4}
  \kappa'(\sigma) F_{\mu\nu}^aF^{\mu\nu}_a + g_0\bar{\Psi}\Psi = 0 \; , \\
  \label{maxwell}
  \partial^{\mu} (\kappa F_{\mu\nu}^a) = j_{\nu}^a \; ,
\eea
where (\ref{dirac}) is the Dirac equation describing the motion of the 
quarks, (\ref{klegor}) is the Klein-Gordon equation for the time evolution
of the $\sigma$-field, and (\ref{maxwell}) are the Maxwell equations for
the gluon fields with the color charge currents
$j_{\nu}^a = -ig_v\bar{\Psi}\gamma_{\nu}{\lambda_a \over 2}\Psi$.
Solutions of these coupled set of nonlinear equations have up to now
only been obtained in
static calculations \cite{FbLe77,BiBiMaWi85,BiBiWi88},
since dynamical simulations suffer from instabilities,
due to the small current quark mass and large Dirac-sea contributions.
Here we will determine the dynamics of the quarks by a transport equation
derived by Elze and Heinz \cite{ElHe89}. These authors have shown on
a semiclassical basis, that the following set of equations for the
phase space distribution functions in the abelian limit can be derived:
\bea
  \label{elze1}
  (p_{\mu}\partial^{\mu}-m^*\partial_{\mu}m^*\partial^{\mu}_p)f^0(x,p) =
  g_vp_{\mu}F^{\mu\nu}_a\partial_{\nu}^pf^a(x,p) \; , \\
  \label{elze2}
  (p_{\mu}\partial^{\mu}-m^*\partial_{\mu}m^*\partial^{\mu}_p)f^a(x,p) =
  g_vp_{\mu}F^{\mu\nu}_a\partial_{\nu}^pf^0(x,p) \; .
\eea
Note that these equations (\ref{elze1}) and (\ref{elze2})
are the transport equations of motion from the Friedberg-Lee Lagrangian,
whereas the corresponding equations (2.14) of \cite{ElHe89} are derived
from the QCD Lagrangian. $f^0$ and $f^a$
\be
  \label{dens}
  f^0(x,p) = \int dQ f(x,p,Q) \;\;\; , \;\;\;
  f^a(x,p) = \int dQ Q^a f(x,p,Q)
\ee
are the {\em color singlet} and the {\em color octet} distribution functions
of the general distribution function $f(x,p,Q)$ for colored particles,
with the space-time vector $x$, the four-momentum $p$ and an
eight component c-number color vector, $Q=Q^a$.
From these distribution functions we obtain the scalar density
$\rho_s = \bar{\Psi}\Psi$, and the color charge density $j_0^a$ needed for the
equations of motion (\ref{klegor}) and (\ref{maxwell}) by
\bea
  \rho_s = {\eta \over (2\pi)^3} \int d^3p {m^* \over \omega}f^0(x,p)
  \;\;\; , \;\;\;
  j_0^a = {\eta \over (2\pi)^3} \int d^3p f^a(x,p) \; ,
\eea
with the effective mass $m^* = m_0 + g_0\sigma$ and the factor
$\eta = 4$ from spin and flavor degeneracy \cite{VeBiMo95}. \\
Because the color fields in (\ref{ftensor}) and (\ref{maxwell}) are
purely abelian the equations (\ref{maxwell}) decouple to eight identical
Maxwell equations. Adopting the Coulomb gauge $\vec{\nabla}(\kappa
\vec{A}) = 0$, these equations can be written as \cite{BiGoWi85}
\bea
  \label{maxsca}
  \vec{\nabla}(\kappa\vec{\nabla}A_0) = -j_0 \; .
\eea
Following the general assumptions of bag models \cite{DeGrJaJoKi75},
we assume that the contributions from the current $\vec{j}^a$ is
already absorbed in the {\em current quark mass} $m_0$. Hence we focus on
the colorelectric field described by (\ref{maxsca}) only, especially in
view of $q\bar{q}$-production from strong colorelectric fields.\\
We make the following particular ansatz for $f^0$ and $f^a$
\bea
  \label{ansatz1}
  f^0(x,p) = f(x,p) + \bar{f}(x,p) \; , \\
  f^1(x,p) = ... = f^7(x,p) = 0 \; , \\
  \label{ansatz3}
  f^8(x,p) = f(x,p) - \bar{f}(x,p) \; .
\eea
This ansatz corresponds to working in a U(1)
subspace of SU(3)-color. This can be seen by remembering that there always
exists an (abelian) gauge transformation such that the color vectors of the 
quarks contribute only to the {\em commutating} generators of SU(3),
$\hat{T}^3$ and $\hat{T}^8$ \cite{SiWe78}. Adopting this expansion,
the quark ($f$) and anti-quark ($\bar{f}$) distribution
functions appearing in (\ref{ansatz1}) and (\ref{ansatz3}) can be
interpreted as quark and {\em di-quark} distribution functions. Nevertheless,
we will in the following refer to ($\bar{f}$) as the anti-quark
distribution function, the U(1) equivalent of the SU(3) di-quark.
%
Inserting the ansatz (\ref{ansatz1})-(\ref{ansatz3}) into the equations for
the color moments (\ref{elze1}) and (\ref{elze2}) we end up with
\bea
  \label{vlas1}
  (p_{\mu}\partial^{\mu}-m^*(\partial_{\mu}m^*)\partial^{\mu}_p)f(x,p) =
  g_vp_{\mu}F^{\mu\nu}\partial_{\nu}^pf(x,p)  \\
  \label{vlas2}
  (p_{\mu}\partial^{\mu}-m^*(\partial_{\mu}m^*)\partial^{\mu}_p)\bar{f}(x,p) =
  -g_vp_{\mu}F^{\mu\nu}\partial_{\nu}^p\bar{f}(x,p) \; ,
\eea
which is a set of usual Vlasov equations describing the motion of charged
particles in a selfconsistently generated scalar and vector field,
determined by equation (\ref{klegor}) and (\ref{maxsca}), respectively.
The coupling of quark and antiquark degrees of freedom is therefore
only provided by their interaction with the mean fields $\sigma$ and
the colorelectric field.\\
Let us first examine the static limit of the transport equations. From quantum
mechanics we know that the color charge density in a hadron has to vanish
locally \cite{moselbuch}
\be
  \label{vanrho}
  \int d^3p (f(x,p)-\bar{f}(x,p)) = j_0^8(x) = <N|\hat{Q}^8(x)|N> = 0 \; ,
\ee
because the nucleon ground state $|N>$ is a color singlet and $\hat{Q}^8$ a color
octet operator. From that we conclude that $f = \bar{f}$ and hence
there is no colorelectric field in the groundstate. As shown by Vetter et. al
\cite{VeBiMo95}, the distribution functions have to be of a local
Thomas Fermi type
\be
  \label{thofer}
  f(x,p) = \bar{f}(x,p) = \Theta (\mu-\omega) \; ,
\ee
with the Fermi energy $\mu$.
After inserting the local Thomas Fermi distributions (\ref{thofer}) into
the integral expressions (\ref{dens})
the fermions can be integrated out, so that we are left with determining the
soliton solution of the remaining $\sigma$-field equation (\ref{klegor}).
This solution is then obtained by standard methods used by Vetter et. al
\cite{VeBiMo95}. Comparing this solution with grounstate properties of
the nucleon, the parameters of the model, $g_0$ and the ones of the
potential $U(\sigma)$, can be fixed.\\
Since the colorelectric field vanishes in the static (ground state) solution
(\ref{vanrho}), the strong coupling constant $\alpha_s = g_v^2 / 4 \pi$
cannot be fixed from ground state properties. In order to determine its value
we investigate the colorelectric energy in the cavity by forming a string
like configuration. The string tension $\tau$, i. e. the coefficient
of the linear rising $q\bar{q}$-potential
\be
  \label{qqbpot}
  V_{q\bar{q}}(r) = - {k \over r} + \tau r \;\;\; , \;\;\; k = {4\alpha_s \over 3}
\ee
within soliton bag models has already been determined by several groups
\cite{BiBiMaWi85,DeGrJaJoKi75,grabiak}. In these studies, the
quark motion is neglected and an instantaneous electric field between them
is assumed ab initio.
We proceed differently and calculate the response of the colorelectric field
to an adiabatic separation of the quark and anti-quark distributions with a constant
velocity. Since we work in the testparticle picture \cite{Wo82},
a fixed initial velocity
$v_z$ is assigned to each particle belonging to $f$ and $-v_z$ to each
particle belonging to $\bar{f}$. The velocity of the quarks
is kept fixed during the time evolution. In this way the source
densities for the time evolution of the $\sigma$-field and the instantaneous
$A^0$-field equations (\ref{klegor}) and (\ref{maxsca}) are obtained.
Several values of $v_z$, ranging from $v_z = 0.1 c$ to $v_z = 0.5 c$, have been
applied to assure the independence of the colorelectric field on $v_z$.\\
At the beginning of the motion we see in fig. \ref{rhosig}
how the cavity is deformed by the motion of the quarks ($t = 3 fm$).
While the charge distributions move apart the colorelectric field, as a
solution of (\ref{maxsca}), builds up (fig. \ref{eglinit}a).
The genuine feature of this field is, that it is parallel everywhere at
the boundary of the cavity, which is due to
the von Neumann boundary conditions, forcing the normal component of the
displacement field $\vec{D} = \kappa \vec{E}$ to vanish. Hence we find
the typical string like behaviour of the $\vec{E}$-field
in the later stages of the time evolution (fig. \ref{eglinit}b), namely
being constant and almost independent of the axial coordinate in the region
between the quarks. Determining the colorelectric energy $E_{glue} =
{1 \over 2} \int d^3r \vec{D} \vec{E}$ as a function of the
$q\bar{q}$-distance provides us with a measure for the string constant $\tau$.
Depending on the
value of $\alpha_s$ the string constant is shown in fig. \ref{tstring}.
In order to obtain the generally accepted value of $\tau \approx 1 GeV/fm$,
we need to have values of $\alpha_s \approx 2$, which is in agreement
with the values of MIT-bag and other Friedberg-Lee model calculations
\cite{BiBiMaWi85,DeGrJaJoKi75,grabiak}.\\
%
Having fixed the strong coupling constant, we now perform the time
evolution of the full model equations for the $\sigma$-field (\ref{klegor}),
the gluon field (\ref{maxsca}) and the transport equations (\ref{vlas1})
- (\ref{vlas2}). As a first example, we excite the soliton ground state at
time $t=0$ by shifting the equilibrium (Thomas Fermi) momentum distribution
(\ref{thofer}) of the quarks with a fixed momentum $+p_z$ and that of the
anti-quarks with $-p_z$. (Values used for $p_z$ range from
$p_z = \mu = 350 MeV$ to $p_z = 30\mu$).
A self-consistent time evolution of the equations of motion (\ref{klegor})
and (\ref{vlas1})-(\ref{vlas2}) is then generated by a numerical integration
sceme as used by Vetter \cite{VeBiMo95}. The instantaneous
colorelectric field is determined in every timestep of the evolution
by a finite element method developed by Mitchell \cite{mitchell}.
The general features of the resulting process can be summarized as follows:\\
Depending on their initial momenta, the quarks move apart along the z-axis
and the electric field builds up. As a consequence, the quarks lose their
kinetic energy to the $\vec{E}$-field and additionally to the deformation
of the cavity to a string like configuration.
When the entire kinetic energy is deposited into the $\vec{E}$- and $\sigma$-field,
the motion is stopped and starts to reverse.
The quarks then pass through each other at the origin
and the process is repeated in the opposite direction, leading to an
oscillation of the $q\bar{q}$-system.
If we follow the process further, the coherent motion of the $q\bar{q}$-system
is disturbed due to an increasing chaotic motion of the testparticles inside
the deformed cavity
\cite{BuBaRa94}. As a consequence, the strength of the colorelectric field
decreases as shown in (fig. \ref{coleng}) for an initial energy of $3 GeV$,
accompanied by a steady elongation of the cavity.
This behaviour continues until the total loss of the colorelectric energy is
used for a large amplitude vibration of the bag (fig. \ref{rms}).
For very large times the superposition of the $q$ and $\bar{q}$
distributions is complete and the colorelectric field vanishes, so that
the final state is again a locally color neutral one as it was
at $t=0$.\\
Note that the elongation of the cavity shown in fig. \ref{rms} is caused by
the initial momentum $p_z$ that acts as an additional pressure on the cavity.
A further effect of the chaotic motion to be expected is the damping of
multipole modes in Vlasov-type transport models \cite{BuBaRa94}, leading to an
equal distribution of the particle momenta to all degrees of freedom.
This damping can hardly be seen in our particular scenarios, since the
excitation energies (typically of the order of a few GeV) are too large, so that
the equipartition of the particle momenta happens on a much larger time scale.
For lower exitation energies of $\approx 300 MeV$, however, this one body
friction effect, driving the bag to a spherical shape, can be seen.\\
Summarizing the statements above, we can say that
independent of the relative momentum of the quarks, the motion is
always restricted to the cavity by the mutual interaction of the
$\vec{E}$-field and the $\sigma$-field with the quarks inside the dynamically
deformed cavity.
Thus, the inherent feature of the
Friedberg-Lee model, color confinement, is maintained in our dynamical
model, i. e. for an arbitrary momentum of the quarks.\\
Further applications of our model will be the effects of color gluon fields
in nucleon-nucleon collisions, similar to our investigations in the
colorless model \cite{KaVeBiMo93,VeBiMo95}, or the description of
quark gluon plasma scenarios, in particular the dynamical treatment of
the hadronization process. Also a consistent description
of $q\bar{q}$-production can be implemented via the Schwinger mechanism.
Work along this direction is in progress.

\newpage
\begin{figure}
\caption{Baryon density $\rho = \int (f + \bar{f}) d^3p$ (left) and the
$\sigma$-field (right) at time $t=3fm$ (upper) and $t=40fm$ (lower).
The equidistant lines of constant value start at $0.1 fm^{-3}$ for the
quark density $\rho$ with an increase of $0.1 fm^{-3}$. The values for
$\sigma$ start at $-0.125 fm^{-1}$ with an increase of $0.05 fm^{-1}$. }
\label{rhosig}
\end{figure}
\begin{figure}
\caption{The electric field $\vec{E}$ inside the cavity. a) At time $t = 3 fm/c$
and b) at time $t=40fm/c$. Note, that b) has a rescaled
z-axis to fit the page.}
\label{eglinit}
\end{figure}
\begin{figure}
\caption{String constant $\tau$ as a function of the strong coupling
constant $\alpha_s$.}
\label{tstring}
\end{figure}
\begin{figure}
\caption{Time evolution of the colorelectric energy up to $t = 100
fm/c$ resulting from an exitation energy of $3 GeV$ (dotted line).
For the solid line the strong oscillations have been averaged out.}
\label{coleng}
\end{figure}
\end{document}